\documentstyle{article}
\def\Msun{\hbox{$\rm\thinspace M_{\odot}$}}

\def\yr{{\rm\thinspace yr}}
\def\h50{\hbox{$\rm\thinspace h_{50}$}}
\def\h50m1{\hbox{$\rm\thinspace h_{50}^{-1}$}}
\def\P3M{\hbox{$P^{3}M$}}

\def\AP3M{\hbox{$AdP^{3}M$}}

\def\cc2{c2}
\def\cc3{c3}
\def\cc4{c4}
\def\cc{c}

\def\ApJ{ApJ}
\def\ApJL{ApJ}

\def\MN{MNRAS}

\def\ref{\par \noindent \hang}

\def\eg{{\it e.g.\ }}

\def\spose#1{\hbox to 0pt{#1\hss}}
\def\approxlt{\mathrel{\spose{\lower 3pt\hbox{$\sim$}}
	\raise 2.0pt\hbox{$<$}}}
\def\approxgt{\mathrel{\spose{\lower 3pt\hbox{$\sim$}}
	\raise 2.0pt\hbox{$>$}}}

\def\<{\thinspace}
\def\boxit#1{\vbox{\hrule\hbox{\vrule\kern3pt\vbox{\kern3pt
          #1 \kern3pt}\kern3pt\vrule}\hrule}}

\begin{document}
\huge
\centerline{\bf Hydra2.0 Documentation}
\vspace{1 cm}
\Large
\centerline{
H.M.P.Couchman\footnote{Department of Physics and Astronomy, 
University of Western Ontario, Ontario, Canada},
F.R.Pearce\footnote{Physics Department, University of Durham, Durham, UK}$^{,3}$,
P.A.Thomas\footnote{Astronomy Centre, University of Sussex, Falmer, Brighton, UK}
}
\vspace{0.5cm}
\centerline{couchman@coho.astro.uwo.ca}
\centerline{F.R.Pearce@durham.ac.uk, P.A.Thomas@sussex.ac.uk} 
\vspace{1cm}
\large
Hydra can be obtained from either of the following websites;

\begin{itemize}
\item http://coho.astro.uwo.ca/pub/hydra/hydra.html

\item http://star-www.maps.susx.ac.uk/\~\/pat/hydra/hydra.html
\end{itemize}

\section{Documentation}

Hydra is an adaptive particle-particle, particle-mesh plus smoothed
particle hydrodynamics N-body simulation program. It can be used with
either periodic or isolated boundary conditions. A compiler flag
allows the gas calculation to be turned off, converting Hydra to a
collisionless mode.

An installation guide comes with the release which provides
information about how to set
up and compile Hydra with parameters appropriate for your particular
problem. This documentation describes the Hydra input files in
more detail.

In what follows code extracts and variables will be written
in {\tt typewriter font}. Program and subroutine names 
will appear in {\tt {\bf bold face}} and directory names
will be bracketed {\tt <thus>}.

\subsection{Data format}

The structure of data files is defined in {\tt {\bf dumpdata.f}}:
\vskip 0.25cm
\begin{verbatim}
      real rm(N),r(3,N),v(3,N),dn(N),e(N),h(N)
      integer itype(N)
      write(8) ibuf,ibuf1,ibuf2
      write(8) rm
      write(8) r
      write(8) v
      write(8) h
      write(8) e
      write(8) itype
      write(8) dn
\end{verbatim}
\vskip 0.25cm

(i) The first line writes out the values of important parameters and
variables as listed in {\tt {\bf pinfo.inc}}:
\vskip 0.25cm
\begin{verbatim}
      common/param/itime,itstop,itdump,itout,time,atime,htime,dtime,
     &         Est,T,Th,U,Radiation,Esum,Rsum,cpu,
     &         tstop,tout,icdump,padding,Tlost,Qlost,Ulost
      common/start/irun,nobj,ngas,ndark,L,intl,nlmx,perr,
     &         dtnorm,sft0,sftmin,sftmax,h100,box100,zmet,spc0,
     &         lcool,rmgas,rmdark,rmnorm,tstart,omega0,xlambda0,h0t0
      common/outputtime/tout1
      real tout1(25)
      dimension ibuf1(100),ibuf2(100),ibuf(200)
      equivalence (ibuf1,itime),(ibuf2,irun),(ibuf,tout1)
\end{verbatim}
\vskip 0.25cm

Each of these variables serves the following function;

\noindent {\tt  itime, itstop, itdump, itout:} step number, stopping step, dumping
interval for backups, not used

\noindent {\tt  time, atime, htime, dtime:} time, expansion factor, hubble parameter,
timestep (all in code units)

\noindent {\tt  Est, T, Th, U:} starting, kinetic, thermal and potential energies in
code units

\noindent {\tt  Radiation, Esum, Rsum, cpu:} radiated energy, energy integral, total
radiated energy, cputime counter

\noindent {\tt  tstop, tout, icdump, padding:} not used, next dump time, position in
list of dump times, padding for isolated boundary conditions

\noindent {\tt  Tlost, Qlost, Ulost:} lost kinetic, thermal and potential energies
in grid units (isolated boundary conditions only)

\noindent {\tt  irun, nobj, ngas, ndark:} run number, total number of particles,
number of gas, dark particles

\noindent {\tt  L, intl, nlmx, perr:} grid size (top level), interlacing switch, max
number of refinement levels, maximum 2-body percentage force error

\noindent {\tt  dtnorm, sft0, sftmin, sftmax:} timestep multiplier, current-day
softening (grid units), min, max softening (grid units)

\noindent {\tt  h100, box100, zmet, sft0:} hubble parameter, boxsize ($h^{-1}Mpc$),
metallicity of gas, average particle spacing

\noindent {\tt  lcool, rmgas, rmdark, rmnorm:} cooling switch, mass of gas,
dark matter particle, force normalisation

\noindent {\tt  tstart, omega0, xlambda0, h0t0:} start time, omega$_0$, lambda$_0$, H$_0$t$_0$
(last three for expanding box only)

\noindent {\tt  tout1:} list of desired output times.

(ii) data arrays 

Note that the following arrays are defined for all particles even
though some of them are irrelevant for dark matter particles.  This is
to simplify book-keeping.  We recommend setting {\tt e}, {\tt h} and
{\tt dn} to zero for dark matter particles.

\vskip 0.25cm
\begin{tabular}{lll}
Array    & used for \\
\hline
{\tt rm} & mass \\
{\tt r}  & positions \\
{\tt v}  & velocities \\
{\tt e}  & thermal energy        & only meaningful for gas \\
{\tt h}  & sph smoothing length  & only meaningful for gas \\
{\tt dn} & density               & only meaningful for gas \\
{\tt itype} & particle type (meaning set in {\tt {\bf itype.inc}}) \\  
\end{tabular}
\vskip 0.25cm

The current settings for {\tt itype} are:

\vskip 0.25cm
\begin{tabular}{cl}
{\tt itype} & used for \\
\hline
-2 & particle does not exist (useful for merging particles) \\
-1 & star (equivalent to dark matter) \\
0 & dark matter \\
1 & gas \\
2 & temporary setting for an already completed gas particle \\
sph evaluation \\
\end{tabular}
\vskip 0.25cm

These can be changed to label particles in different ways but you
should check that Hydra handles the resultant types correctly.

\subsection{Units}

WARNING: the units used internally by Hydra differ from those used in
the datafiles.  The reason for this is that internally Hydra uses
positions in the interval [1,L+1), where L is the number of FFT grid
cells across the box.  The data is either mapped into this interval
exactly or, for isolated boxes and refinements, a padding region is
left around the outside of the box.  L can be different in different
refinements and can vary from run to run - it has no physical meaning
but is purely a computational device.
The units also differ in isolated and periodic simulations:

(i) isolated simulations (usually compiled with -DISOLATED).
\begin{itemize}
\item  length: positions run from 0 to 1 in each co-ordinate
direction.  The length-unit is {\tt box100} $h^{-1}Mpc$.
\item mass: currently set to $10^{10} \Msun$.  Can be altered.
\item time: currently set to $10^{10} \yr$.  Can be altered.
\item speed: length/time
\item density: number of particles per box volume.  Converted
internally to a number density of ions + electrons using {\tt nunit}.
\item temperature: stored in units of internal energy (ie
speed$^2$).  Convert to ergs by multiplying by {\tt eunit = vunit$^2$};
convert to temperature by multiplying by {\tt Kunit = eunit*2.*mum/3./kb}
\end{itemize}

(ii) cosmological simulations (periodic boxes)
\begin{itemize}
\item length: positions run from 0 to 1 in each co-ordinate
direction.  The length-unit is {\tt atime*box100} $h^{-1}Mpc$.
\item mass: the total mass in the box is normalised to match the
desired value of $\Omega_0$.  Thus particle masses are all relative.
\item time: normalised to unity at the present, ie {\tt tunit}=1 at
the end. $t_0$ is calculated automatically from input cosmological 
parameters.
\item speed: The output velocities are dx/dt where x is the spatial
coordinate in the range [0,1). The peculiar velocity is thus {\tt atime *
dx/dt} and the proper velocity {\tt datime/dt*x + atime*dx/dt}. 
\item density: number of particles per box volume.  Converted
internally to a number density of ions + electrons using nunit.
\item temperature: stored in units of internal energy (ie
speed$^2$).  Convert to ergs by multiplying by {\tt atime$^2$*eunit = vunit$^2$};
convert to temperature by multiplying by {\tt Kunit = atime$^2$*eunit*2.*mum/3./kb}
\end{itemize}

The units are initialised in the routine {\tt {\bf inunit.F}} to which you
should refer if in doubt.  They are also written at the head of the
log file.

The magnitude of the force-scaling depends upon the system of units
which is in use.  It is set by the factor rmnorm defined at the head
of {\tt {\bf updaterv.F}}.

\subsection{Creating initial conditions}

The input data arrays which need to be defined for all particle
species are {\tt rm, r, v} and {\tt itype}.  In addition gas particles need an
initial temperature, {\tt e}, and sph smoothing length, {\tt h}.  This latter
quantity may be estimated as the mean interparticle separation,
{\tt dn$^{-1./3.}$}).  {\tt dn} itself is not required.

For dark matter runs the following are the minimum set of input
parameters which need to be defined: {\tt itime}, {\tt time},{\tt
irun}, {\tt nobj}, {\tt h100},
{\tt box100}, {\tt omega0} (periodic boxes only), {\tt xlambda0} (ditto), {\tt tout1}.
In addition for gas runs the cooling flag needs to be set: {\tt lcool}=1 for
cooling, 0 otherwise, and for {\tt lcool}=1 then the metallicity, {\tt zmet},
should be defined in terms of the Solar value.

An example initial conditions generator for an isolated box is
{\tt {\bf createtop.f}}, included with this distribution.  A periodic initial
conditions generator is also included, the {\tt {\bf cosmic}} package.

\subsection{Running Hydra}

Hydra works within and below a runtime directory. An example of the
required directory setup was automatically built by unpacking the tar
file. The executable ({\tt {\bf hydra}}) will be copied to the run directory
{\tt <rundir>} automatically when it is made (the destination directory can
be reset by changing RUNDIR in the makefile).  When Hydra is executed
the runlog ({\tt {\bf hydra $>$ runlog}}) 
and a summary file ({\tt {\bf pr????.log}} where {\tt {\bf ????}}
is the run number) are also written out here whilst the data is
written to (and read from) a sub-directory {\tt <data>}.  The Green's
functions the code uses to do its PM calculation are saved to the
sub-directory {\tt <greenfn>}. 
If you have more than one {\tt <rundir>} then it is usually best
to link together the greenfn directories by using {\tt ln -s ...}.
So the directory tree looks like this;
\vskip 0.25cm
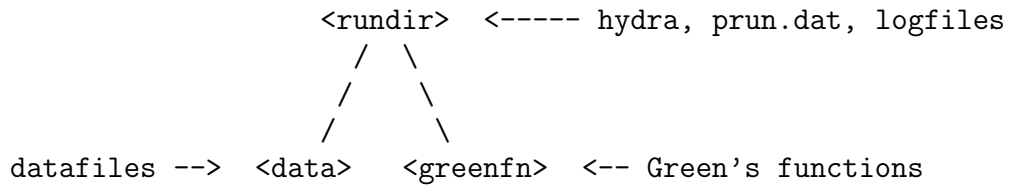
\begin{verbatim}
                        <rundir>  <----- hydra, prun.dat, logfiles
                          /  \
                         /    \
                        /      \
     datafiles -->  <data>   <greenfn>  <-- Green's functions
\end{verbatim}
\vskip 0.25cm

\noindent Hydra uses a short file to set its basic run parameters. 
This file, {\tt <rundir>/{\bf prun.dat}} has the following format:

\vskip 0.25cm
\begin{tabular}{ll}
d3712.0857         &   startup data file name \\
3724               &  irun; run number for this run \\
858 10             &   itstop itdump \\
1.0                &   dtnorm \\
0.02               &   sft(Mpc/h) \\ 
1                  &   refinements on/off \\
\end{tabular}
\vskip 0.25cm

The first line contains the name of the datafile you wish to start
with. This is built up of {\tt {\bf d????.nnnn}} where {\tt {\bf ????}} is a run number and
{\tt {\bf nnnn}} is the step number. You can reset the run number on the second
line. Here we are starting with the 857th step of run 3712, calling
this run 3724.  The third line contains the step number you wish to
stop at and how frequently you require incremental backup files.
dtnorm is the timestep normalisation: this should normally be set to
unity (see Section 3.8 below).  Next comes the Plummer softening in
units of $h^{-1}Mpc$ (see Section 3.7 below).  The final line contains a
switch for turning refinement placing on or off. You can prevent Hydra
placing refinements and convert it into a P3M code by setting this
parameter to 0.

\subsection{Log files}

Hydra sends output to the screen and to a summary file.  The screen
output can be redirected to a log file (\eg {\tt {\bf nohup hydra
$>$ run????.log }})
but it is quite large and useful mainly for diagnostic purposes.

The summary file is called {\tt {\bf pr????.log}} where {\tt {\bf ????}}
is the run number.
One line is written to the summary file every step.  Its form differs
slightly between periodic and isolated runs:

(i) periodic boxes

\vskip 0.25cm
\begin{tabular}{ccccccc}
step & cputime & time & redshift & $K$ & $U$ \\
857 & 927.48 & 0.5440 & 0.5005 & 5.00E+07 & 2.52E+06 \\
858 & 966.93 & 0.5445 & 0.4997 & 5.01E+07 & 2.52E+06 \\
& & & $W$      & error & $(K+U)/a$ & $W/a$ \\
& & & 7.41E+07 & 0.0065 & 7.89E+07 & 1.11E+08 \\
& & & 7.42E+07 & 0.0065 & 7.88E+07 & 1.11E+08 \\
\end{tabular}
\vskip 0.25cm

where $K$= Kinetic energy, $U$= Thermal energy, $W$= Potential energy
and $a$= {\tt atime}.
The final two columns should be in the ratio 2:3 (approximately -
ignoring softening) and should remain constant in time during the
linear regime.  The error is defined in terms of the ratio of the
change in the Layzer-Irvine energy integral, $I$, to the potential
energy.  $I$ is defined as $I=K+U+W+\int[2(K+U)+W]da/a+\int Ldt$ where
$L$=radiated power and $t=time$.  See Couchman, Thomas \& Pearce
(1995) for details. 

(ii) isolated boxes

\vskip 0.25cm
\begin{tabular}{cccccccc}
step & cputime & time & K & U & W \\
0 & 112.58 & 0.0000 & 0.00E+00 & 0.00E+00 & 1.26E+01 \\
1 & 112.25 & 0.0438 & 6.12E-03 & 3.20E-10 & 1.26E+01 \\
& & & error   & lost K & lost U & lost W \\
& & & 0.0000  & 0.00E+00 & 0.00E+00 & 0.00E+00 \\
& & & 0.0005  & 0.00E+00 & 0.00E+00 & 0.00E+00 \\
\end{tabular}
\vskip 0.25cm

The error estimate in this case is $[K+U+W-lost(K+U+W)-starting(K+U+W)]/U$.

The screen output begins with some warning messages informing the user
if any parameters in the input data file have been altered.  It also
prints the size, L, of the top-level grid and confirms the presence or
absence of refinements.  Next follows a list of units (remember to
include appropriate atime factors in expanding boxes).
Next comes some diagnostic information:

\vskip 0.25cm
\begin{verbatim}
 itime, time, atime, htime: 1295    1.000    1.000    0.667
 refinement:   0,  scaling factor:  1.00
  ndark, ngas, ndone, nstar:     32768    32768        0        0
  noveredge, novergrid:      2578     3697
  gravity: av,min,max # neighbours    110.22         0      1864
      sph: av,min,max # neighbours     23.48         0       161
	   av,min,max density (ref.)        7.6971        0.0139      383.5992
	   av,min,max  energy (base)       50.3896        0.0000    45135.1523
 refinement:   1,  scaling factor:  3.29
  ndark, ngas, ndone, nstar:      2247      906     1241        0
  noveredge, novergrid:       213       94
  gravity: av,min,max # neighbours     41.12         0       391
      sph: av,min,max # neighbours     43.19        28        96
	   av,min,max density (ref.)        1.0347        0.0239        7.8099
	   av,min,max  energy (base)      124.5373        0.0739      433.4357
\end{verbatim}
\vskip 0.25cm

 ...similarly for other refinements...

\noindent{\tt timestep =   2.6515502E-04 (2.6515502E-04  7.3740509E-04  9.3750297E-02)}
\vskip 0.25cm

(a) `scaling factor' is the expansion factor of the refinement grid relative
to the base level.  

(b) ndark, ngas, ndone and nstar give the number of particles of each
type within that refinement: ndone refers to those gas particles whose
sph forces have already been calculated at the previous level. 

(c) The average minimum and maximum number of gravity neighbours
refers to the pp force calculation: the average value should 
be about 100 for efficient distribution of work between PM and PP.

(d) The minimum number of sph neighbours can drop below 32 in
low-density regions because we only search for neighbours out to a
radius of approximately 2.2 times the grid-spacing: within refinements the
minimum number of sph neighbours should be close to 32.  The maximum
number of sph neighbours can be larger because the smoothing length
can never drop below the gravitational softening.

(e) The average, minimum and maximum density of the gas particles is
given in refinement units (ie particles per refinement grid cell).  The
average, minimum and maximum temperature of gas particles is scaled
back to the base level units.  These units both differ from those in
the data files.  These entries are omitted from the log if there are
no gas particles within that refinement.

(f) The final line gives the timestep and the constraint on the
timestep from accelerations, velocities and the hubble expansion (see
Section 3.8 below).

\subsection{Softening}

Hydra uses softening shape function which has compact support:
\vskip 0.25cm
\begin{verbatim}
      SUBROUTINE S2(r,a,grav,dpot)
      xi=2.*r/a
      if(xi.ge.0. .and. xi.lt.1.)then
       grav=xi*(224.+xi*xi*(-224.+xi*(70.+xi*(48.-xi*21.))))/35./a**2
       dpot=(208.+xi*xi*(-112.+xi*xi*(56.+xi*(-14.
     &                           +xi*(-8.+xi*3.)))))/70./a
      else if(xi.ge.1. .and. xi.lt.2.)then
       grav=(12./xi**2-224.+xi*(896.+xi*(-840.+xi*(224.+xi*(70.+
     &        xi*(-48.+xi*7.))))))/35./a**2
       dpot=(12./xi+128.+xi*(224.+xi*(-448.+xi*(280.
     &        +xi*(-56.+xi*(-14.+xi*(8.-xi)))))))/70./a
      else 
       grav=1./r**2
       dpot=r*grav
      end if
      RETURN
      end 
\end{verbatim}
\vskip 0.25cm

It is approximately equivalent to a Plummer softening with extent
$a/2.34$ (the central value of r/force is the same in each case).  The
user supplied value for the softening in prun.dat is assumed to be a
Plummer softening and is therefore multiplied by 2.34 before
converting to grid units.
If the ISOLATED or COMOVING flag is set during compilation then the
softening is held fixed.  Otherwise it scales with time as 
{\tt $soft=min(0.6,sft0/atime)$} where {\tt $sft0$} is the user-supplied value and the
maximum value of 0.6 is a numerical constraint.

When choosing a softening it is important to strike a balance between
the competing desires of high spatial resolution and long timesteps.
dtime scales approximately in proportion to $t_s=\sqrt{(soft^{3}/Gm)}$ where
m is the mass of the highest-mass objects in the simulation.  One
should also be aware of the danger of artificial two-body relaxation
which will occur in high-density regions on a timescale
$t_2=\sqrt{45N}t_s$ where $N$ is the number of particles within the
softening.  See Thomas \& Couchman (1992) for more more a detailed discussion.

\subsection{Timestepping}

Our choice of timestepping algorithm is described in Couchman, Thomas
\& Pearce (1995).  We
evaluated several algorithms and settled for a simple PEC scheme which
utilises only the current positions (plus velocities and temperatures
for gas particles) to determine the forces.  This has the advantage of
keeping storage to a minimum and also allowing arbitrary changes in
timestep if the force changes abruptly in time (as it can do in the
vicinity of shocks).  The scheme is equivalent to Leapfrog (but less
memory-efficient) if there are no gas forces.

The timestep used by Hydra is not spatially variable but it does vary
from one step to the next.  The timestep is set by examining the
greatest acceleration and velocity present at the current time:

\vskip 0.25cm
\begin{verbatim}
      dta=(soft2/asqmax)**0.25
      dtv=sqrt(soft2/vsqmax)
      dtime=min(0.25*dta,0.4*dtv)
      if (htime.gt.1e-30) then
        dtime=min(dtime,0.0625/htime)*dtnorm
      else
        dtime=min(dtime,1.)*dtnorm
      end if
\end{verbatim}
\vskip 0.25cm

\noindent where {\tt soft2} is the softening length squared, {\tt asqmax} is the square of
the maximum acceleration and {\tt vsqmax} is the square of the maximum
velocity.  For expanding boxes the {\tt htime} condition ensures that
cooling due to the Hubble expansion is followed accurately: this
dominates at early times.

\subsection{SPH parameters}

The sph smoothing length is chosen so as to encompass approximately 32
neighbours.  Some people prefer a higher value but we have found that
this makes an imperceptible difference to the results (32 particles
gives significant shot-noise scatter in the interpolated density for a
Poisson distribution but the actual particle distribution is much more
uniform than this).  In low-density regions the maximum sph search
length (which is limited to approximately 2.2 times the grid-spacing)
encloses fewer neighbours; in high-density regions the minimum sph
search-length (which is set equal to the S2 softening parameter ---
see Section 3.7 above) may enclose more particles.
If a particle's sph search length either extends beyond
the refinement boundary or is more than 2.2 grid-spacings in
size then the sph force for that particle is calculated at
the previous refinement level (resulting in a loss of efficiency).
The number of particles for which this is done is given in the log.

We use a compact smoothing kernel:
\vskip 0.25cm
\begin{verbatim}
      wnorm=1./4./pi
      if (x.le.1.0) then
        kernel=wnorm*(4.-6.*x**2+3.*x**3)
      else if (x.le.2.0) then
        kernel=wnorm*(2.-x)**3
      end if
\end{verbatim}
\vskip 0.25cm

\noindent where {\tt x=r/h} is the ratio of the particle separation to the smoothing
length, except that the gradient is modified so as to ensure a
force which decreases monotonically with radius (see {\tt {\bf kernel.f}} and
Thomas \& Couchman 1992).

The artificial viscosity is based on the bulk divergence of the flow,
not on the relative pairwise velocity of particles.  Schematically:

\vskip 0.25cm
\begin{verbatim}
      frc=twothird*e
      if (divpph.lt.0.) frc=frc+(bvisc*divpph-avisc*cs)*divpph
\end{verbatim}
\vskip 0.25cm

\noindent where {\tt divpph=h*div(v)}, {\tt cs} is the sound speed and the
parameters {\tt avisc}
and {\tt bvisc} are set to 1 and 2, respectively.  The actual code, which
includes a correction for the hubble expansion, is buried in the heart
of {\tt {\bf shgravsph.F}}.

\subsection{Cooling}

Cooling can be turned on or off by setting the parameter {\tt lcool}
(0=off, 1=on).  The supplied cooling function is a simple series of
power-law fits to the optically-thin radiative cooling code of
Raymond, Cox \& Smith:

\vskip 0.25cm
\begin{verbatim}
      if (t.lt.1e5) then
        Lambda=(1.4e-28+1.7e-27*zmet)*t
      else
        Lambda=5.2e-28*t**0.5+1.4e-18*t**(-1.)+1.7e-18*t**(-0.8)*zmet
      end if
\end{verbatim}
\vskip 0.25cm

\noindent where {\tt t} is the temperature in Kelvin.  This fit contains contributions
from both bremsstrahlung and recombination cooling from hydrogen and
helium plus a variable contribution from metal lines with an assumed
abundance of {\tt zmet} times solar.  The fit is good to within about
a factor of two above $10^4K$ --- below this temperature the cooling
function drops precipitously and we take it to be zero.  We intend to
incorporate a more accurate cooling function in a future release.

Because the cooling time is often shorter than the dynamical time, we
do not use it to limit the timestep.  Instead we allow the particles
to evolve adiabatically, then cool them, assuming a constant density,
at the end of the timestep.

\subsection{Brief description of the code}

\begin{verbatim}
MAIN:
   startup: read in parameters and data
   inunit: define the units
   loop until finished:
   updaterv: PEC step
   |    accel: acceleration including hubble drag; timestep evaluation
   |        force: acceleration evaluation
   |           rfinit: initialise refinements
   |           refforce: see below
   |           clist: create particle lists for refinements
   |           loop over refinements:
   |           load: load particles into refinement
   |           |  refforce: see below
   |           |  uload: unload particles from refinements
   |           end loop
   |     infout: write summary file
   |  output: write out data and backup files
   end loop
   end

refforce:

 green     : evaluate, or read, green's function
 mesh      : evaluate PM accelerations
 list      : sort particles into search cells
 refine    : determine the position of subrefinements
 shforce   : tabulate PP force and potential
 shgravsph : apply PP and SPH forces; 
             write out diagnostic information
\end{verbatim}
\vskip 0.25cm

\subsection{Cosmological initial conditions}

A cosmological initial conditions generator -- {\tt {\bf cosmic}}
comes with the Hydra package. It produces an initial conditions file
in Hydra format from a supplied data file, {\tt {\bf cosmic.dat}}.
There are in act 3 executables in the cosmic package. These are
{\tt {\bf cosmic}} itself which takes {\tt {\bf cosmic.dat}}
for input and outputs the Hydra datafile into {\tt <data>{\bf
d????.0000}}, where {\tt {\bf ????}} is {\tt irun} from 
{\tt {\bf cosmic.dat}} and {\tt {\bf r.pert}} is a perturbation
file read by {\tt {\bf peakfindfft}}. {\tt {\bf peakfindfft}}
searches {\tt {\bf r.pert}} for peaks of a given size
(the default is $8h^{-1}Mpc$) and outputs the position and $\delta$ of
the biggest peak into {\tt {\bf peak.dat}}. Finally there is {\tt {\bf
pnsum}} which compares the sums of waves in boxes to the true 
power spectrum in spheres of radius $8h^{-1}Mpc$ 
(only useful if boxsize exceeds this) using the parameter file
{\tt {\bf cosmic.dat}}.

\noindent{\tt {\bf cosmic.dat}} has the following format;

\vskip 0.25cm
\begin{verbatim}
1000    -1066601        irun,iseed1
4096                    ndark 
20      0.5             box100,h100
1.      0.              omega0,xlambda0
.06     0.              omegab0,gamma
2       0               ispec,ind
0.02    0.6             atime,sigma8
0       0.5             lcool,zmet
\end{verbatim}
\vskip 0.25cm
Notes;
\begin{itemize}
 \item  {\tt ndark} must be the cube of a power-of-2
 \item  {\tt omegab=0} for dark matter only,
        {\tt omegab$>0$} gives {\tt ndark} each of dm and gas with correct mass ratio
 \item if {\tt gamma$<=0$} then uses formula in Viana \& Liddle (1996)
 \item 
\begin{itemize}
\item	{\tt ipsec} =     1 - power law (index {\tt ind})
\item	{\tt ipsec} =	    2 - cdm
\item	{\tt ipsec} =	    3 - hdm
\end{itemize}
 \item {\tt sigma8} is normalised to initial time using formulae in Viana \&
 Liddle (1996)
 \item {\tt lcool=1} for cooling (0 for no cooling), {\tt zmet} is in solar units
\end{itemize}

\subsection{Utility programs}

\subsubsection{readheader}

The program {\tt {\bf readheader}} tells you some useful
information about your initial conditions (or any dataset). It will
automatically be placed into {\tt <rundir>} when made.  {\tt {\bf
readheader}} prompts
you for the name of a datafile and checks to see if you have set $Nmax$
and $Lmax$ correctly and also shows the ranges which your data spans if
you request further information. This utility is very useful for
checking data consistency.

\subsubsection{hydra2tipsy}

{\tt {\bf hydra2tipsy}} converts Hydra output files to {\tt {\bf
tipsy}} format. 
It converts positions to $h^{-1}Mpc$, velocities to $km/s$, temperatures to
Kelvin and densities to overdensities (or atoms per $cm^3$ if an
isolated box has been asked for). {\tt {\bf hydra2tipsy}} is also automatically
placed into {\tt <rundir>} by make and prompts for a Hydra datafile. 
The output file is in the format {\tt {\bf tip????.xxxx}} where {\tt
{\bf ????}} is the
run number and {\tt {\bf xxxx}} is the current step.
Tipsy, an N-body visualisation package can be obtained from;
\begin{itemize}
\item http://www-hpcc.astro.washington.edu/tools/TIPSY/
\end{itemize}

\section*{Acknowledgments}

PAT held a Nuffield Foundation Science Research Fellowship during 1995/96.
FRP was an EPSRC PDRA working for the Virgo Consortium. 
HMPC is supported by NSERC of Canada.
We acknowledge a NATO (CRG 920182) travel grant which facilitated our
interaction.

\section*{References}

\noindent Couchman, H. M. P., 1991, \ApJL, 368, L23

\noindent Couchman, H. M. P., Thomas, P. A., Pearce, F. R., 1995,
\ApJ, 452, 797

\noindent Pearce, F. R., Couchman, H. M. P., 1997, astro-ph/9703183

\noindent Raymond, J. C., Cox, D. P., Smith, B. W., 1976, \ApJ, 204, 290

\noindent Thomas, P. A., Couchman, H. M. P., 1992, \MN, 257, 11

\noindent Viana, P. T. P., Liddle, A. R., 1996, \MN, 281, 323

\end{document}